\documentclass[twocolumn,twoside,slac_two]{revtex4}
\usepackage{graphicx}
\usepackage{fancyhdr}
\usepackage{multirow}
\begin{document}

\title{Mesons From String Theory}

%

\author{K. Stiffler}
\affiliation{Department of Physics and Astronomy, The University of Iowa, Iowa City, IA 52242, USA}

\begin{abstract}
A brief historical synopsis of the connection between gauge theories and string theory is given. 
Meson configurations known as $k$-strings are examined from string theory via the gauge/gravity correspondence.  Backgrounds dual to $k$-strings in both $2+1$ and $3+1$ are discussed.  The energy of $k$-strings to lowest order consists of a tension term, proportional to the length, $L$, of the $k$-string, i.e., the size of the mesons in the configuration.  The first quantum correction is a Coulombic $1/L$ correction, known as a L\"uscher term, plus a constant.  Acquiring tensions and L\"uscher terms via the gauge/gravity correspondence is discussed.
\end{abstract}

\maketitle

\thispagestyle{fancy}


\section{Introduction}From its inception, string theory has long thought to have a deep connection with the strong nuclear force.  From a series of calculations by t' Hooft, Green, Schwarz, Maldacena, Herzog, Klebanov, et. al.~\cite{'tHooft:1973jz,Green:1984sg,Green:1984fu,Green:1984ed,Maldacena:1997re,Herzog:2001fq},  today we see the connection through the gauge/gravity correspondence.  As string theory can describe gauge theories in the case of open strings, and supergravity (SUGRA) in the case of closed strings, we can hope to find a map from SUGRA to gauge theories, known as the gauge/gravity correspondence.  The specific case we will investigate is performing calculations done near the horizon of a SUGRA solution, and relating it to a strongly coupled gauge theory calculation.  

The stongly coupled gauge theory example we use is the $k$-string, which we find particular SUGRA solutions to be dual to in either $2+1$ or $3+1$ dimensions.  The $k$-string is an assemblage of fundamental strings, where the fundamental string is a quark and an anti-quark source connected by a color flux tube of length $L$, as described in~\cite{Luscher:1980ac}.  For large quark separations, $L$, the energy of $k$-strings is dominated by the tension term, a term proportional to $L$.  The lowest order correction is a Coulombic term, proportional to $1/L$.  Both of these terms can be found from supergravity duals of $k$-strings.  For $2+1$ $k$-strings, we will demonstrate this specifically in the background of Cvetic, Gibbons, Lu, and Pope (CGLP)~\cite{Cvetic:2001ma}. We will compare this tension to preliminary results for another background dual to $2+1$ $k$-strings:
the background of Maldacena and Nastase (MNa)~\cite{Maldacena:2001pb}.  We will also show results of $3+1$ $k$-string calculations from the background of Klebanov and Strassler (KS)~\cite{Klebanov:2000hb}
\section{SU($N$) $k$-strings}

\subsection{L\"uscher's Fundamental String}

\begin{figure}[h]
\centering
\includegraphics[width=80mm]{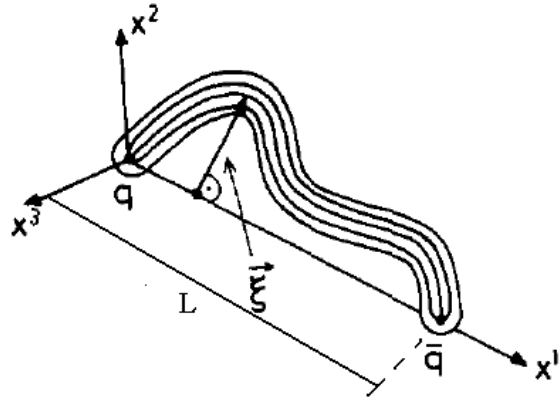}
\caption{L\"uscher's fundamental string: a color flux tube forms between a quark and an anti-quark separated by a distance $L$.~\cite{Luscher:1980ac}} \label{fig:LuscherStringNoCaption}
\end{figure}

L\"uscher's picture of the fundamental string is as in Figure~\ref{fig:LuscherStringNoCaption}.  L\"uscher found the energy of this configuration to be of the form
\begin{equation}\label{eq:fstringenergy}
   E = T L + \beta - \frac{\pi(d-2)}{24 L} + \mathcal{O}(L^{-2}),
\end{equation}

\noindent where $T$ is known as the tension, and $d$ is the dimension of spacetime. The $1/L$ correction term is the so-called, \emph{L\"uscher term}.

.

\subsection{The $k$-string}
\begin{figure}[h]
\centering
\includegraphics[width=80mm]{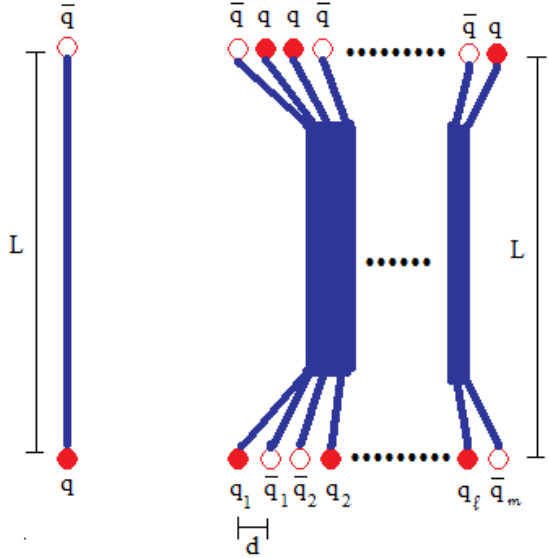}
\caption{The fundamental string(right) and a $k$-string(left).} \label{fig:kstrings}
\end{figure}

Here is a quick overview of $k$-strings.  A more complete review can be found in~\cite{Shifman:2005eb}.  SU($N$) $k$-strings are an assemblage of fundamental strings, a close distance $d << L$ next to each other as in Figure~\ref{fig:kstrings}, where $k = |l - m|$, $l$  being the number of quarks on one side of the $k$-string and $m$ being the number of anti-quarks on that same side.

For large $L$, $k$-strings exhibit $k$-ality: the tension vanishes whenever $k = N$.  This $k$-ality is exhibited in models from lattice gauge theory~\cite{Lucini:2001nv,Bringoltz:2006zg,Bringoltz:2008nd}, Hamiltonian methods~\cite{Karabali:1998yq,Nair:2003em,Karabali:2007mr}, and supergravity calculations using the gauge/gravity correspondence~\cite{Herzog:2001fq,Herzog:2002ss}.  Two possible forms of the $k$-string tension, $T_k$, are
\begin{equation}\label{eq:sinecasimir}
   T_k \propto \left\{ \begin{array}{l l}
                     N \sin\frac{k \pi}{N} & \mbox{sine law} \\
                     k \frac{N - k}{N} & \mbox{casimir law}
                \end{array}
                \right.
\end{equation}

\noindent where clearly either law exhibits $k$-ality.  Table~\ref{tab:comparetensions} compares $k$-strings tensions in $2+1$ dimensions calculated from various methods.  From this data, it would seem that in $2+1$ dimensions, the casimir law is more appropriate.  It also shows that the supergravity calculations in $2+1$ done here more closely align with the anti-symmetric quark representations.

\begin{table}[h!]
\centering
\caption{Comparison of $k$-string tensions from various methods.  The values quoted are $T_k/T_f$, where $T_k$ is the $k$-string tension, and $T_f$ is the fundamental string tension, i.e., $k=1$.  The CGLP tension is calculated from the transcendental Eqs.(\ref{eq:Hmin},\ref{eq:constraint}); MNa(Sine) and Casimir from Eq.(\ref{eq:sinecasimir}). Data in quark representations:
S=symmetric~\cite{Karabali:2007mr}, A=antisymmetric~\cite{Karabali:2007mr}, M=mixed~\cite{Karabali:2007mr}, *=antisymmetric~\cite{Bringoltz:2008nd}}
\label{tab:comparetensions}
{\footnotesize
\begin{tabular}{|c|c|c|c|c|c|c|}
\hline
$Group$ & $k$ &  CGLP & MNa(Sine) & Casimir & lattice & Karabali-Nair  \\
\hline\hline
\multirow{2}{*}{$SU(4)$} & \multirow{2}{*}{2}  & \multirow{2}{*}{1.310} & \multirow{2}{*}{1.414}  & \multirow{2}{*}{1.333} &  1.353(A) & 1.332(A)   \\
&&&&& 2.139(S) & 2.400(S)\\
\hline\hline
$SU(5)$ & 2 &  1.466 & 1.618 & 1.5 & 1.528* & 1.529*  \\
\hline\hline
\multirow{5}{*}{$SU(6)$} & \multirow{2}{*}{2} & \multirow{2}{*}{1.562} & \multirow{2}{*}{1.732} & \multirow{2}{*}{1.6} & 1.617(A) & 1.601(A) \\
\cline{6-7}
&&&&& 2.190(S) & 2.286(S)\\
\cline{2-7}
& \multirow{3}{*}{3} & \multirow{3}{*}{1.744} & \multirow{3}{*}{2.0} & \multirow{3}{*}{1.8} & 1.808(A) & 1.800(A) \\
\cline{6-7}
&&&&&3.721(S) & 3.859(S) \\
\cline{6-7}
&&&&&2.710(M) & 2.830(M)\\
\hline\hline
\multirow{3}{*}{$SU(8)$} & 2 & 1.674 & 1.848 & 1.714 & 1.752* & 1.741*\\
\cline{2-7}
&3 & 2.060 & 2.414 & 2.143 & 2.174* & 2.177*\\
\cline{2-7}
&4 & 2.194 & 2.613 & 2.286 & 2.366* & 2.322*\\
\hline
\end{tabular}
}
\end{table}

Furthermore, lattice calculations done by Bringholtz and Teper~\cite{Bringoltz:2008nd} find a L\"uscher term in $2+1$ dimensional $SU(5)$ and $SU(4)$ gauge theory to be close to 
\begin{equation}
    -\frac{\pi}{6L},
\end{equation}

\noindent the same value we calculated with the CGLP SUGRA model in~\cite{Doran:2009pp} which is dual to a $2+1$ $SU(N)$ $k$-string configuration.  At first glance, it may be troubling that these results are different from L\"uscher's, as seen in Eq.~\ref{eq:fstringenergy}.  However, L\"uscher's result was for the fundamental string, which we do not expect to exhibit the precise behavior of a $k$-string, which is a series of fundamental strings "glued together", as in Fig.~\ref{fig:kstrings}.
\section{$K$-strings from Supergravity Dual Theories}
\begin{figure}[h]
\centering
\includegraphics[width=80mm]{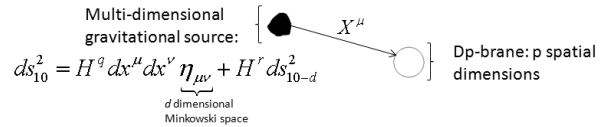}
\caption{A probe Dp-brane embedded in a SUGRA background.} \label{fig:embedding}
\end{figure}

Through the gauge gravity correspondence, we expect a string theory embedded in a SUGRA background to be dual to a gauge theory with a large number, $N$, of colors. Investigating $k$-string dual SUGRA solutions, Herzog and Klebanov~\cite{Herzog:2001fq,Herzog:2002ss} considered an embedding as in Figure~\ref{fig:embedding}, where a probe Dp-brane, either electrically($Q$) or magnetically($M$) charged
\begin{equation}
   F = dA = Q dt \wedge dx + M d\theta \wedge d\phi,
\end{equation}

\noindent is embedded in a classical SUGRA background, typically of the form
\begin{equation}\label{eq:typicalbackground}
ds_{10}^2 = H^q\eta_{\mu\nu}dx^{\mu}dx^{\nu} +H^p ds_{10-d}^2,
\end{equation}

\noindent sourced by
\begin{eqnarray}
   F_{n+1}(X^{\mu}) &=& d C_n(X^{\mu}), ~~~\Phi(X^{\mu}), \nonumber\\
   H_3(X^{\mu}) &=& d B_2(X^{\mu}),
\end{eqnarray}

\noindent where $t,x,\theta,\phi$ are four of the Dp-brane coordinates, $\zeta^a$, and $X^{\mu}$ are the bosonic SUGRA coordinates.   It is important to note here that $d$ is the spacetime dimension of the Minkowski spacetime portion of the metric in Eq.~\ref{eq:typicalbackground}, which will be the spacetime dimension in which the $k$-string will be embedded in the gauge dual theory.

The SUGRA coordinates become fields on the Dp-brane, the field theory dynamics governed by the Dp-brane action
\begin{eqnarray}
  S_{p}& = & -\mu_p \int d^{p+1}\zeta e^{-\Phi}\sqrt{-\det(g_{ab} + \mathcal{F}_{ab})} + \nonumber\\
    & &~~~~~~~~~+\mu_p \int \sum_n C_n\wedge \mathcal{F} + S_f
\end{eqnarray}

\noindent where
\begin{equation}
   \mathcal{F}_{ab} = B_{ab} + 2\pi \alpha' F_{ab},~~~\mu_p = (2 \pi)^p(\alpha')^{(p+1)/2}
\end{equation}

\noindent and $S_f$, which is a functional solely of fermionic fields, $\Theta$, on the Dp-brane, is classically set to zero~\cite{PandoZayas:2008hw,Doran:2009pp}.  Considering classical solutions ($A_0$, $X_0$) where the only field with dynamics is the electric field component of $F$,
\begin{equation}
    S_{p} = \int \mathcal{L}(A_0,\dot{A}_0,X_0)
\end{equation} 

\noindent we can apply the Legendre transformation to the Dp-brane action, yielding the Hamiltonian:
\begin{equation}
   \mathcal{H} = \frac{\partial \mathcal{L}}{\partial \dot{A}} \dot{A} - \mathcal{L} 
\end{equation}

\noindent Minimization of this Hamiltonian leads to the $k$-string tension~\cite{Herzog:2001fq,Herzog:2002ss}
\begin{equation}
  \mathcal{H}_{min} = T_k L
\end{equation}

The first quantum corrections are found by fluctuating around the classical solution
\begin{equation}\label{eq:fluctuate}
   X^{\mu} = X^{\mu}_0 + \delta X^{\mu},A^{\mu} = A^{\mu}_0 + \delta A^{\mu},\Theta = 0 + \delta \Theta,
\end{equation}

\noindent expanding out the action to second order in these fluctuations
\begin{equation}\label{eq:expandaction}
  S_{p} = S_{p}^{(0)} + S_{p}^{(1)} + S_{p}^{(2)},
\end{equation}

\noindent and calculating the free energy of the one loop corrections through
\begin{equation}\label{eq:partitionfunction}
   e^{E_1 T} = Z_2 = \int DX DA D\bar{\Theta} D\Theta e^{iS_{p}^{(2)}}.
\end{equation}

Notice the one loop energy is found from the quantum physics of the second order action, $S_{p}^{(2)}$, as the first order action $S_{p}^{(1)}$ vanishes when evaluated at the classical field equations, as it should.  The one loop energy found this way through the gauge/gravity correspondence is~\cite{PandoZayas:2008hw,Doran:2009pp}
\begin{equation}\label{eq:Luscherterm}
  E_1^{(d,p)} = -\frac{\pi(d + p - 3)}{24 L} + \beta_d
\end{equation}

\noindent where $\beta_d$ is constant with respect to $L$.  

\section{$K$-string from CGLP Supergravity Background}
Following the outline of the previous section, we summarize the work of~\cite{Herzog:2002ss,Doran:2009pp} where the $2+1$ dimensional $k$-string energy was calculated as the dual of a D4-brane embedded in the CGLP background.  
\subsection{The CGLP Supergravity Background}
First, we briefly review the CGLP SUGRA background.  The complete details can be found in the original paper~\cite{Cvetic:2001ma}.  The CGLP background is a type IIA supergravity background, sourced by
\begin{eqnarray}
  F_4 &=& g_s^{-1}d^3x \wedge d H^{-1} + m(f_4 \epsilon_{ijk}\mu^idr \wedge J^k +\nonumber\\
      &&~~~~~~~~~~+ f_5 X_2 \wedge J_2 + f_6 J_2 \wedge J_2)  \\
 l H_3 &=& f_1 dr\wedge X_2 + f_2 dr\wedge J_2 + f_3 X_3 \\
  e^{\Phi} &=& g_s H^{1/4}
\end{eqnarray}

\noindent with all other type IIA supergravity sources set to zero.

In the above, we have
\begin{eqnarray}
   X_2 & = &\frac{1}{2}\epsilon_{ijk}\mu^i D\mu^i \wedge D\mu^k,~~~J_2 = \mu^i J^i \\
  D\mu^i & = & d\mu^i + \epsilon^{ijk}A^j \mu^k \\
  J^i &=& dA^i + \frac{1}{2}\epsilon^{ijk}A^j\wedge A^k.
\end{eqnarray}

\noindent and $J^i$ satisfies the algebra of unit quaternions.  With these sources, the background takes the form
\begin{eqnarray}
   ds^2 & = & H^{-1/4}dx^{\mu}dx^{\nu}\eta_{\mu\nu} + H^{1/4}l^2[h^2 dr^2 + \nonumber\\
       &&~~~~~~~~~~~~~~~~+ a^2(D\mu^i)^2+ b^2 d\Omega_4^2]
\end{eqnarray}

\noindent where the bosonic supergravity coordinates are the set
\begin{equation}
   X^{\mu} = (x^0, x^1, x^2, r, \mu^1, \mu^2, \mu^3, \psi,\chi,\theta,\phi)
\end{equation}

\noindent with the constraint $(\mu^i)^2 = 1$.  In the above, $H, h, a, b,$ and $f_i$ are functions of r, and $m$ and $l$ are consants. In fact
\begin{equation}
   m = 8 \pi \alpha'^{3/2}g_s N
\end{equation}

\noindent where $N$ is the number of parallel D4-branes sourcing the background and is also the number of colors in the gauge theory dual.
\subsection{SU($N$) $K$-string Tension and L\"uscher Term from CGLP Background}
Here we show a brief outline of the full calculation of the tension and L\"uscher term, which can be found in~\cite{Doran:2009pp}.  We use a probe D4-brane embedded in the CGLP SUGRA background.  As the Minkowski spacetime poriton of the CGLP background is $2+1$ dimensional, the dual $SU(N)$ $k$-string will be in $2+1$ dimensions.

The classical action for a probe D4-brane in the CGLP background is
\begin{eqnarray}
   S^{(0)} & =  -\mu_4 \int d^5\zeta e^{-\Phi}&\sqrt{-det(g_{ab} + 2\pi\alpha' F_{ab}} + \nonumber\\
     &&+\mu_4\int C_3 \wedge F
\end{eqnarray}

\noindent where $F$ is electrically charged. Constructing the Hamiltonian and minimizing with respect to the bosonic SUGRA coordinate $\psi$ yields
\begin{eqnarray}\label{eq:Hmin}
  \mathcal{H}_{min} &=& \alpha N L \sin^2\psi_0\sqrt{\sin^2\psi_0 + (3\alpha/q)^2\cos^2\psi_0} \nonumber\\
 &=& T_k L
\end{eqnarray}
\noindent subject to the constraint
\begin{equation}\label{eq:constraint}
   \frac{4 k}{3N} = \xi(\psi_0) + 3 \frac{\alpha^2}{q^2}\sin^2\psi_0\cos\psi_0
\end{equation}

\noindent with
\begin{equation}
   \xi(\psi_0) = \int_0^{\psi_0}\sin^3u du
\end{equation}

\noindent and where $\alpha$ and $q$ are constants with $\alpha/q \approx 0.3083$ and $\psi_0$ is the classical value of $\psi$ whose solution is the solution to the constaint Eq.~\ref{eq:constraint}.  

\begin{figure}[h]
\centering
\includegraphics[width=80mm]{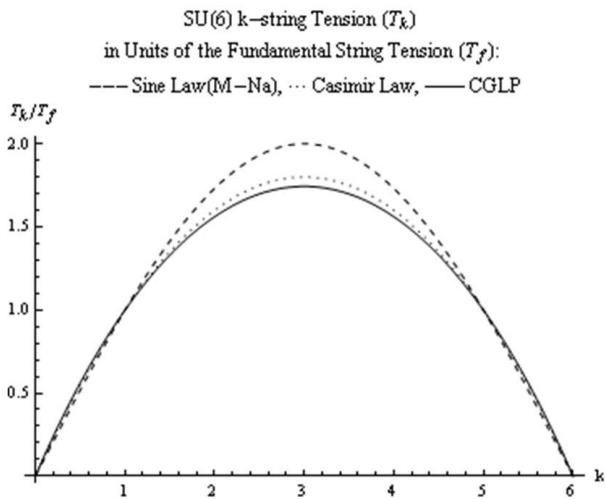}
\caption{Plot of CGLP $k$-string tension with sine law and Casimir law for $N$ = 6 colors.  Plots look similar for increasing $N$.} \label{fig:kstringplot}
\end{figure}

When we fluctuate about this classical solution by the method outlined in Eqs.~\ref{eq:fluctuate},~\ref{eq:expandaction}, and~\ref{eq:partitionfunction}, we find the one loop energy to be
\begin{equation}
   E_1 = -\frac{\pi}{6L} + \beta_3
\end{equation}

\noindent which contains a term constant of $L$, $\beta_3$, which arises from the massive modes, plus a L\"uscher term from the massless modes, $-\pi/6L$, which is the same as that found by lattice calculations of Bringholtz and Teper~\cite{Bringoltz:2008nd}.  We find we can group this new calculation of the L\"uscher term together with a previous calculation dual to $3+1$ $k$-strings\footnote{Here we correct for a factor of 1/2 missing in~\cite{PandoZayas:2008hw}}, into a single formula
\begin{equation}\label{eq:CGLPLuscherterm}
    V_{\mbox{L\"uscher}} = -\frac{(d + p - 3)}{24 L}
\end{equation}

\noindent where $d$ is the dimension of the Minkowski spacetime portion of the SUGRA and also the dimension in which the dual $k$-string lives and $p$ is the spatial dimension of the probe Dp-brane.
\section{Conclusion}
We have given a brief summary of the gauge/gravity correspondence and shown the method for calculating $k$-string tensions and L\"uscher terms from the SUGRA side of this correspondence.  The tension of $k$-strings in $2+1$ calculated from the CGLP background seems to align well with anti-symmetric quark representations, as shown in Table~\ref{tab:comparetensions}.  The L\"uscher term found from the CGLP dual theory, Eq.~\ref{eq:CGLPLuscherterm}, aligns well with Lattice calculations of Bringholtz and Teper~\cite{Bringoltz:2008nd}.  Furthermore, our current findings for the L\"uscher terms found in $2+1$ and $3+1$ can succinctly be written as in Eq.~\ref{eq:CGLPLuscherterm}.

\begin{acknowledgments}
I would like to thank my advisor, Vincent Rodgers, and my other collaborators, Leopoldo A. Pando-Zayas and Christopher A. Doran, for their work done in~\cite{PandoZayas:2008hw,Doran:2009pp}, off which this proceeding is based.  I would also like to thank the organizers of the DPF 2009 conference~\cite{organizers}.  This work is partially supported by Department
of Energy under grant DE-FG02-95ER40899 to the University of Michigan and the National
Science Foundation under award PHY - 0652983 to the University of Iowa.
\end{acknowledgments}
\bigskip 

\end{document}